\let\oldvec\vec% Store \vec in \oldvec
\documentclass{llncs}
\let\newvec\vec% Store publisher's \vec in \newvec
\let\vec\oldvec% Restore \vec from \oldvec so amsmath will not complain when trying to change it

\usepackage{amsmath}
\let\vec\newvec% Restore \vec as set by publisher

\PassOptionsToPackage{usenames,dvipsnames,svgnames,table}{xcolor}

\usepackage[utf8]{inputenc}
\usepackage[english]{babel}
\usepackage{silence}
\usepackage[labelfont=bf,labelseparator=period]{caption}

\usepackage{subcaption}
\usepackage{cite}
\usepackage{todonotes}
\usepackage{graphicx}

\usepackage{xcolor}

\graphicspath{{}{./fig/}{./source/}}

\usepackage[utf8]{inputenc}

\usepackage[final,linktocpage]{hyperref}

\usepackage{cleveref}
\crefname{figure}{Figure\kern-0.3em}{figure\kern-0.3em}
\crefname{table}{Table\kern-0.3em}{table\kern-0.3em}

\begin{document}

\title{Tools for Analyzing Parallel I/O}%: \\ A Comprehensive Survey
\author{Julian M. Kunkel,$^1$ Eugen Betke,$^2$  Matt Bryson,$^3$  Philip Carns,$^4$ Rosemary Francis,$^5$  Wolfgang Frings,$^6$ Roland Laifer,$^7$   Sandra Mendez$^8$}
\institute{$1$: University of Reading, $2$: German Climate Computing Center (DKRZ), \\ $3$: University of California, Santa Cruz,  $4$: Argonne National Laboratory, $5$: Ellexus Ltd,   $6$: J\"ulich Supercomputing Centre (JSC), $7$: Karlsruhe Institute of Technology (KIT),  $8$: Leibniz Supercomputing Centre (LRZ) }
\date{\today}
\maketitle

\begin{abstract}
Parallel application I/O performance often does not meet user expectations.
Additionally, slight access pattern modifications may lead to significant changes in performance due to complex interactions between hardware and software.
These issues call for sophisticated tools to capture, analyze, understand, and tune application I/O.

In this paper, we highlight advances in monitoring tools to help address these issues.
We also describe best practices, identify issues in measurement and analysis, and provide practical approaches to translate parallel I/O analysis into actionable outcomes for users, facility operators, and researchers.
\end{abstract}

\section{Introduction}
\label{sec:introduction}
The efficient use of I/O systems is of prime interest for data-intensive
applications, since storage systems are increasing in size and the use case of a single system is so diverse, especially in the scientific community.
As computing centers grown in size, and the high-performance computing (HPC) community approaches exascale
\cite{bensonunderstanding2009, bergman2008exascale}, it has become increasingly important to understand how these systems are operating and how they are being used.
Additionally, understanding system behavior helps light the path for future storage system development and allows the purveyors of these systems to ensure performance is adequate to allow for work to continue unimpeded.
While industry systems are typically well understood, shared storage systems in the HPC community are not as well understood.
The reason is their sheer size and the concurrent usage by many users, which typically submit disparate workloads.
Most applications achieve only a fraction of theoretically available performance.
Hence, optimization and tuning of available knobs for hardware and software
are important.
This requires that the user to understand the I/O behavior of the
interaction between application and system, since it determines the runtime of the applications.

However, measuring and assessing observed performance are already
non-trivial tasks and raise various challenges around hardware, software, deployment, and management.
This paper describes the current state of the practice with respect to such tools.

\medskip

The paper is structured as follows:
First, an introduction to concepts in performance analysis is given in \Cref{sec:background}.
Various tools and concepts are described in \Cref{sec:tools}.
In \Cref{sec:examples} small example studies illustrate how these tools could be used to identify inefficient system behavior.
In \Cref{sec:issues} we discuss common issues and sketch potential roads to overcome these issues.
In \Cref{sec:conclusion} we summarize our conclusions.

\section{Introduction to Performance Analysis}
\label{sec:background}

In computer science, \textit{performance analysis} refers to activity that fosters understanding in timing and resource utilization of applications.
Understanding resource utilization includes understanding runtime behavior of the application and the system.
For parallel applications, the concurrent computation, communication, and parallel I/O increase the complexity of the analysis.

In industry, the process of system and application tuning is often referred to as \textit{performance engineering}.
Software engineers design special methods to embed performance engineering into the application development.
With these approaches, performance is considered explicitly during the application design and its implementation.
Tools and methodologies such as \textit{Computer-aided software engineering}
(CASE)  serve the developer in the phases of the software life-cycle, namely, requirement engineering, analysis, design, coding, documentation, and testing.
Such tools try to encourage software engineers to incorporate the performance relevant aspects early in the development cycle.
Unfortunately, the research and processes in industry are not integrated in
state-of-the-art HPC application development, although a few tools do assist in the development of parallel applications.
Instead, a closed loop of performance tuning is applied that optimizes programs after a running version exists.
%Since later discussed tools focus on the monitoring and analysis of behavior, we will discuss the measurement and analysis phase in more detail.

So far, we discussed the situation from the point of view of the user or
application. For facility operators or file system administrators, the I/O
performance of the whole system or of one parallel file system is the
primary concern: Does the system provide the expected or degraded
performance, is the system is efficiently used, or do some applications
create such a high load that other applications see a significant
performance impact?

\subsection{Closed Loop of Performance Tuning}

The localization of a performance issue on an existing system is a process in which a hypothesis is supported by measurement and theoretic considerations.
Measurement is performed by executing the program while monitoring runtime behavior of the application and the system.
In general, tuning is not limited to source code; it can be applied to any system.
The typical iterative optimization process is the \textit{closed loop of performance tuning} consisting of the phases:
 \paragraph{\textbf{Measurement}} of performance in an experiment.
The environment, consisting of hardware and software including their
configuration, is chosen; and the appropriate input (i.e., problem
statement) is decided.
%While theoretic considerations allow projecting runtime behavior of arbitrary input sets,
Since monitoring is limited to instances of program input, optimizations
made for a particular configuration might degrade performance on a different setup or program input.
Typically, the measurement itself perturbs the system slightly by degrading performance.
Picking the appropriate measurement tools and granularity can reveal the relevant behavior of the system.

\paragraph{\textbf{Analysis}} of obtained empirical performance data to \textbf{identify issues} and optimization potential in the source code and on the system.
Particularly, hot spots---code regions where execution requires significant
portions of runtime (or system resources)---are identified.
Then, the optimization potential of the hot spots is assessed based on potential performance gains and the estimated time required to modify the current solution.

Changing a few code lines to improve runtime by 5\% is more efficient than
recoding the whole input/output of a program, especially if I/O might account for only 1\% of the total runtime.
However, care must be taken when the potential is assessed; depending on the overall runtime, a small improvement might be valuable.
From the view of the computing facility, decreasing by 1\% the runtime of a
program which runs for millions of CPU hours yields a clear benefit by
saving operational costs in form of 10,000 CPU hours (about 1.5 CPU years).

\paragraph{\textbf{Generation of alternatives}.} Based on the insight gained
by the analysis, alternative implementation and tuning options are explored,
and system modifications are considered that may mitigate the observed performance issue.
This is actually the hardest part of the tuning because it requires that the behavior of the new system can be predicted or estimated.
In practice, however, multiple potential options often are evaluated; and,
based on the results, the best one is chosen.
With increasing experience and knowledge of the person tuning the system,
the number of options is reduced as the future behavior can be better anticipated.

\paragraph{\textbf{Implementation}.} At the end of a loop the current system
is modified; that is, one of the performance relevant layers is adjusted, to realize the potential improvement of the new design.
The system is re-evaluated in the next cycle until the time required to change the system outweighs potential improvements or potential gains are too small because the performance measured is already near-optimal.
In practice, however, in most cases the efficiency of the current solution is not estimated; instead, the current runtime is considered to be potentially saved.

\medskip

These phases are then repeated until the desired performance is achieved or
until further tuning is not cost effective.

\subsection{Measurement}
In the closed loop, data is collected that characterizes the application run and system utilization.
Many types of data can be collected.
For example, the operating system provides a rich set of interesting characteristics such as memory, network, I/O, and CPU usage.
These characterize the activity of the whole system, and sometimes usage can even be assigned to individual applications.

The semantics of this data can be of various kinds. Usually, a \textit{metric} defines the measurement process and the way subsequent values are obtained.
For example, \textit{time} is a simple metric that indicates the amount of time spent in a program, function or hardware command.
The \textit{throughput} of a storage system, that is, the amount of data transferred per second, is another metric.

One way of managing performance information is to store \textit{statistics},
for example, absolute values such as number of function invocations,
utilization of a component, average execution time of a function, or
floating-point operations performed.
Statistics of the activity of a program are referred to as a \textbf{profile}.
A profile aggregates events by a given metric, for example by summing up the inclusive duration of function calls.
Many tools exist that generate profiles for applications.

In contrast to a profile, a \textbf{trace} records events of a program
together with a timestamp. Thus, it provides the exact execution chronology and allows analysis of temporal dependencies.
External metrics such as hardware performance can be integrated into traces as well.
Tracing of behavior produces much more data, potentially degrading
performance and distorting attempts of the user to analyze observation.
Therefore, in an initial analysis, often only profiles are recorded.
A combination of both approaches can be applied to reduce the overhead while
still offering enough information for the analysis.
Events that happen during a timespan can be recorded periodically as a
profile for an interval, thus allowing analysis of temporal variability.
By generating profiles for disjoint code regions, behavior of the different program phases can be assessed.
Another approach is to enable tracing conditionally, for example by
capturing the coarse-grained I/O behavior of the application with statistics
and starting tracing when observed performance drops.

The performance data must be correlated with the interesting application's behavior and source code.
Depending on the measurement process, assigning information to the cause can
be impossible. For example, a statistic cannot reveal the contribution of concurrent activity.
Some low-level tools exploit compiler-generated debug symbols to localize the origin of triggered events in the source code.

\bigskip

Several approaches can be used to measure the performance of a given application.
A \textit{monitor} is a system that collects data about the program execution.
Approaches can be classified based on \textit{where}, \textit{when} and \textit{how} runtime behavior is monitored.
A monitor might be capable of recording activities within an application
(e.g., function calls), across used libraries, activities within the operating system such as interrupts, or it may track hardware activities; in principle, data can be collected from all components or layers.
For I/O analysis, monitors rely on software to measure the state of the
system. Data from available hardware sensors is usually queried from the software on demand.
Hardware monitors are too expensive, complicated, and inflexible to capture program activity in detail, yet some metrics such as network errors may be provided by the hardware itself.

\bigskip

File system administrators typically measure on the file system server side.
Many measurements are independent from the file system type, for example,
the load on the servers or the summarized throughput on network adapters or
on storage subsystems. Some statistics, however, are only available with special file system types (e.g., Lustre, Spectrum Scale, BeeGFS, NFS).

\subsection{Preparation of Applications}

Although the Linux operating system offers various statistics in the \texttt{/proc} file system, this information is often insufficient for I/O analysis.
Usually, changes are made to the program under inspection in order to increase analysis capabilities; the activity that alters a  program is called \textit{instrumentation}.
Popular methods are to modify source code, to relink object files with patched functions, overriding dynamic library calls at execution time with \emph{LD\_PRELOAD}, or to modify machine code directly~\cite{Shende01}.
During execution, such a modified program invokes functions of the monitoring environment to provide additional information about the program execution and the system state.
This instrumentation functionality could also be supported directly by the
(operating) system, and hence one could collect performance data without modifying the application.

Since a software monitor requires certain resources to perform its duty (those can be considered as overhead), monitoring an application perturbs the original execution.
Observed data must be kept in memory and might be flushed to disk if memory space does not suffice.
Additionally, computation is required to update the performance data.
The overhead depends on the characteristics of the application and system: it might perturb behavior of the instrumented application so much that an assessment of the original behavior is impossible.
In I/O analysis, particularly storing the profile or trace in memory and
flushing it to secondary storage incurs considerable overhead that causes additional I/O.

Several techniques can be used to combat potential overhead (and thus
application perturbation) in I/O instrumentation.  Some tools constrain
their instrumentation to summary statistics~\cite{carns200924} or
compressed representations~\cite{vijayakumar2009scalable} to minimize
overhead.  Others may automatically filter activity if an event is
fired too often or if the overhead of the measurement system itself
grows too high.
If filtering still incurs too much overhead, then interesting functions can be manually instrumented, i.e., by inserting calls to the monitoring interface by hand.

Additionally, a selective activation of the monitor can significantly reduce the amount of recorded data.
A monitor could sample events at a lower frequency, reducing the overhead and the trace detail level on the same extent.

\subsection{Analysis of Data}
Users analyze the data recorded by the monitoring system in order to localize optimization potential.
Performance data can be recorded during program execution and assessed after
the application has finished; this approach of \textit{post-mortem} analysis is also referred to as \textit{offline} analysis.
An advantage of this approach is that data can be analyzed multiple times and compared with older results.
Another approach is to gather and assess data \textit{online}, while the program runs.
In this approach, feedback is provided immediately to the user, who could adjust settings to the monitoring environment depending on the results.

Because of the vast amount of data, sophisticated tools are required in
order to localize performance issues of the system, correlate them with
application behavior, and identify the source code causing them. Tools
operate either manually (i.e., the user must inspect the data) or
automatically. A \textit{semi-automatic tool}  could give hints to the user
where abnormalities or inefficiencies are found. Tool environments that localize and tune code automatically, without user interaction, are on the wishlist of all programmers.
Because of the system and application complexity, however, such tools are only applicable for a very small set of problems.
Usually, tools offer analysis capability in several \textit{views} or \textit{displays}, each relevant to a particular type of analysis.

At best a system-wide monitoring of all applications can be conducted to
reveal issues; in other words, all applications running on a supercomputer are constantly monitored in a non-intrusive fashion while additional analysis is triggered upon demand.
In order to better understand the behavior of a single application, separate
analysis runs may be conducted, since it is important to reduce the
complexity of scientific software in order to find the cause of the behavior.

The performance analysis is usually done in an ad hoc manner because of the
nature of the logs produced by these large scale machines: they are machine
specific, and despite having similar attributes, do not share a single
format for their trace behavior. This situation extends past HPC storage,
and applies more broadly to HPC in general and, additionally, to individual application traces.

Because of this difference in log/trace format, system analysis is usually
done per system. Such analysis can produce useful results on the behavior of
a particular system, but sheds no light on how it compares to other similar
systems, HPC or otherwise \cite{adams:sc12 ,adams:tos12, wang:mss04,
grawinkel_analysis_2015}. Additionally, programmer effort is wasted
analyzing each system when the generated analytics are the same or similar
for each system. Because of the difference in traces and techniques, these are not typically comparable, leading to most systems being analyzed in a vacuum, never compared to one another.

\section{Tools}
\label{sec:tools}
This section gives an overview of existing tools.
%An overview of the characteristics of the tools is given in \Cref{tbl:overview}.
%
%\begin{table}
%  \begin{tabular}{ll}
%    \hline
%  \end{tabular}
%  \caption{Overview of existing tools}
%  \label{tbl:overview}
%\end{table}

\subsection{Darshan}
Darshan \cite{carns2011understanding-toc,hpcdarshan} is an open source I/O characterization tool for post mortem analysis of HPC applications' I/O behavior.
Its primary objective is to capture concise but useful information with minimal overhead. Darshan accomplishes this by eschewing end-to-end
tracing in favor of compact statistics such as elapsed time, access sizes, access
patterns, and file names for each file opened by an application.  These statistics are captured in a bounded amount of memory per
process as the application executes.  When the application shuts down, it is reduced, compressed, and stored in a unified log
file.  Utilities included with Darshan can then be used to analyze, visualize, and summarize the Darshan log information.
Because of Darshan's low overhead, it is suitable for system-wide deployment on large-scale systems.  In this deployment model,
Darshan can be used not just to investigate the I/O behavior of individual applications but also to capture a broad view of system
workloads for use by facility operators and I/O researchers.
Darshan is compatible with a wide range of HPC systems.

Darshan supports several types of instrumentation via software modules.  Each module provides its own statistical counters and
function wrappers while sharing a common infrastructure for reduction, compression, and storage.  The most full-featured modules provide
instrumentation for POSIX, MPI-IO and standard I/O library function calls, while additional modules provide limited PNetCDF and HDF5
instrumentation.  Other modules collect system information, such as Blue Gene runtime system parameters or Lustre file system
striping parameters.  The Darshan eXtended Tracing (DXT) module can be
enabled at runtime to increase fidelity by recording a complete
trace of all MPI-IO and POSIX I/O operations.

Darshan uses \emph{LD\_PRELOAD} to intercept I/O calls at runtime in
dynamically linked executables and link-time wrappers to intercept I/O calls
at compile time in statically linked executables. For example, to override
POSIX I/O calls, the GNU C Library is overloaded so that Darshan can
intercept all the read, write and metadata operations. In order to measure
MPI I/O, the MPI libaries must be similarly overridden. This technique allows an application to be traced without modification and with reasonably low overhead.

\subsection{Vampir}
Vampir\footnote{\url{http://www.paratools.com/Vampir}} \cite{Knuepfer2012} is an open source graphical tool for post mortem performance analysis of parallel systems.
It supports off-line analysis of parallel software (MPI, OpenMP,
multi-threaded) and hardware-accelerated (CUDA and OpenCL) applications.
The analysis engine allows a scalable and efficient processing of large amounts of data.
Vampir uses the infrastructure of Score-P\footnote{\url{http://www.vi-hps.org/projects/score-p/}} for instrumenting applications.
Score-P stores events in a file, which can be analysed by Vampir and
converted to different views; for example, events can be presented on a time axis or compressed to different statistics.
Some views have elaborate filters and zoom functions that can provide an overview but can also show details.
Effective usage of Vampir requires a deep understanding of parallel programming.
Although the program enables one to capture and analyze sequences of POSIX I/O operations, it gives little or no information about the origin or evaluation of I/O.

\subsection{Mistral/Breeze}
Mistral is commercial command-line tool from Ellexus\footnote{\url{https://www.ellexus.com/products/}} used to report and resolve I/O performance issues of misbehaving complex Linux applications on HPC clusters.
It has real-time monitoring abilities and can change the I/O behavior of applications in order to delay I/O operations and prevent overloading of shared storage.
Rules for monitoring and throttling I/O are stored as plain text in configuration files called contracts, which can be modified
at runtime by privileged users for systemwide changes and by users
application-wide.
A sophisticated logging mechanism registers monitoring and throttling events
and stores them in files; but with an appropriate
plug-in, the logging information can be redirected to any location, for
example, in a central database such as Elasticsearch or InfluxDB so that the
results can be viewed with Grafana.\footnote{\url{https://grafana.com/}}
Administrators can use Mistral to identify applications that are running with bad I/O patterns and harming performance of shared
file systems. It can also be used outside of production to run quality assurance tests on complex applications prior to deployment. The key idea behind Mistral is that the information collected is configurable and easily aggregated so that Mistral can be run at scale. Mistral supports POSIX and MPI (MPICH, MVAPICH, OpenMPI) I/O interfaces.

Ellexus Breeze is a user-friendly, self-explained, and well-documented off-line analysis tool with command-line, GUI, and HTML reporting modes.
All the information gathered during the application runtime is presented in a comprehensive format and can be of great help to developers.
The detailed information about environment can help support teams reproduce and understand problems.
The most valuable piece of information is the list of application dependencies
so that users and administrators can get a list of every file, library,
program, and network location used by an application.
Breeze also includes a breakdown
of how each program accessed each file so that performance issues such as inefficient metadata access can be found and resolved.

The analysis tool is delivered with a tool called \emph{``trace-program.sh"}
that can capture MPI-IO and POSIX function calls and information about the environment and store them in binary trace files.
It uses \emph{LD\_PRELOAD} to wrap original I/O function and, therefore, works only with dynamically linked I/O libraries.
Breeze uses either a proprietary binary trace format or plain text output files reported similar to the popular diagnostic tool \emph{strace}.
The binary format has some performance advantages and can be decoded to a human readable representation by the included \emph{decode-trace.sh} tool.

Another feature is the ability to compare two different application runs,
thereby identifying changes in an application’s behaviour and providing valuable feedback to the application's developers.

\subsection{SIOX}
SIOX \cite{siox_arch} is a highly modular instrumentation, analysis, and profiling framework.
It contains an instrumentation tool \emph{``siox-inst”}, a trace reader
\emph{``siox-trace-reader”}, and a set of plug-ins and wrappers.

It has wrappers for MPI, POSIX, NetCDF, and HDF5 interfaces that
contain reimplementations of the original I/O functions.
Inside a reimplemented function is a call to the original function or syscall, and instrumentation code, that generates an activity after each execution.
Activities in SIOX are structures that contain various information about the calls.
The wrappers can be dynamically linked to an application by using \emph{LD\_PRELOAD} and the creation of wrappers during link-time.

Extreme modular design is a key feature of SIOX.
The tools \emph{siox-inst} and \emph{siox-trace-reader} can be considered as pure plug-in infrastructures.
In other words, there is no functionality inside until some plug-ins and wrappers are loaded.
Usage of different sets of plug-ins and wrappers may result in ``new” tools
that exactly fit the problem.
There is no restriction on the number of wrappers and plug-ins that can be loaded simultaneously, so that the functionality of SIOX can be easily extended to perform complex tasks.
It has been used to research various aspects such as triggering tracing
depending on unusual system behavior, the creation of replayers for recorded behavior, the mutation of I/O calls rerouting targets or replacing system calls, and online analysis.
Online analysis can be done by \emph{siox-inst}, by collecting activities from the wrappers and forwarding them to the registered plug-ins.
Off-line analysis is based on both tools.
Most of the SIOX plug-ins use plug-in interfaces that are supported by \emph{siox-inst} and \emph{siox-trace-reader}, and consequentially these plug-ins can be used by both tools.
SIOX has also been coupled with OpenTSDB and Grafana to support online
monitoring~\cite{RIOHAWSEGA17}.

By using an instrumented version of FUSE, we found that the complexity of
I/O tracers can be reduced and one can monitor \texttt{mmap()} of applications.
Note that because of the popularity of other tools, SIOX is primarily maintained as a research vehicle.
%%%%%%%%%%%%%%%%%%%%%%%%%%%%%%% BEGIN PIOM-MP %%%%%%%%%%%%%%%%%%%%%%%%%%%%%%%%%%%%%%%%%%%
\subsection{PIOM-MP}
PIOM-MP (formerly known as PAS2P-IO \cite{Mendez:2012,MendezCACIC:2012})
represents the MPI application's I/O behavior by using I/O phases. A phase is a
consecutive sequence of similar access patterns into the logical view of a file.
Because HPC scientific applications show a repetitive behavior,
\texttt{m} phases will exist in the application. PIOM-MP identifies applications’ phases
with their access patterns and weights.
By using phases of the applications, the analysis focuses on the functional model of the applications.

\begin{figure}[bt!]
\vspace{-4mm}
	\centering
	{\hspace*{-4mm}\includegraphics[width=0.9\textwidth]{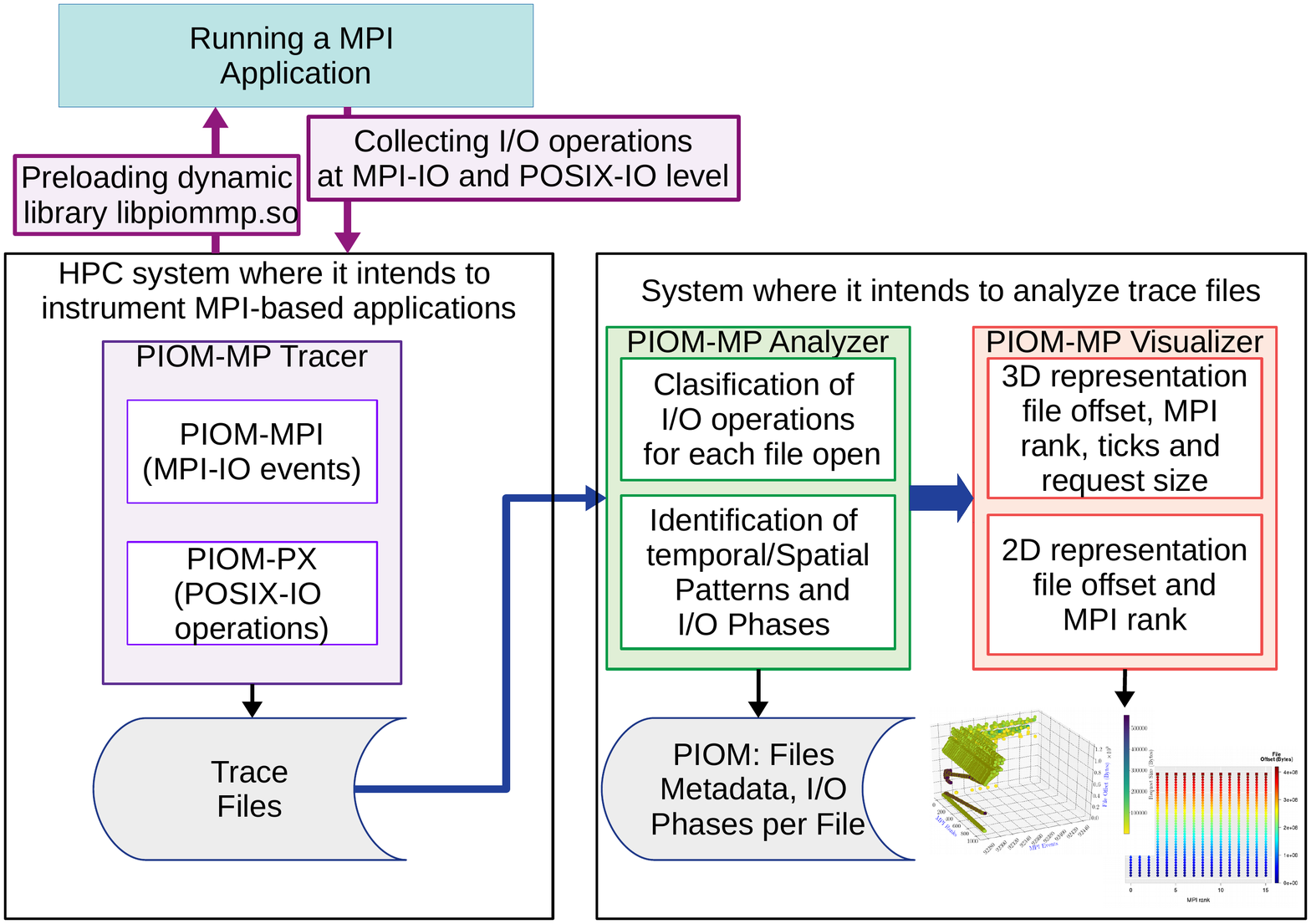}}
	\vspace{-3mm}
	\caption{PIOM-MP modules}
	\label{fig:piommp-mod}
\vspace{-4mm}
\end{figure}

The I/O phases are used as units for analyzing the performance and scalability of
parallel applications.
Our approach\,\cite{GomezSanchez:2017} depicts the global access pattern
(spatial and temporal) at the \texttt{MPI-IO}
and \texttt{POSIX-IO} level.
\Cref{fig:piommp-mod} shows different components of \texttt{PIOM-MP}. It
comprises three modules: \texttt{PIOM-MP} Tracer, \texttt{PIOM-MP} Analyzer, and \texttt{PIOM-MP}
Visualizer. The first module must be in the HPC system in which the parallel application
is executed; the other modules can be in a different system.

With PIOM-MP, the user can define the relation between the phase's pattern
and the I/O system configuration by using the concept of the I/O requirement in order
to explain the I/O performance in a specific system~\cite{Mendez:2017}.
Furthermore, the user can extrapolate the phases for another number of MPI
processes and workloads to evaluate the parallel application I/O scalability.
%%%%%%%%%%%%%%%%%%%%%%%%%%%%%%% END PIOM-MP %%%%%%%%%%%%%%%%%%%%%%%%%%%%%%%%%%%%%%%%%

\subsection{Additional User-Level Tools}

The Linux kernel monitors various statistics and provides them in the \texttt{/proc} file system.
These statistics are updated with a certain frequency---typically 1
second---and can be queried by various Linux tools to monitor system and application behavior.
Since these counters are incremented, they can be used to derive certain statistics for any interval such as the amount of data accessed during the execution of an application.
For example, \emph{collectl}\footnote{collectl, \url{http://collectl.sourceforge.net/}} can be used to capture and analyse Lustre throughput or metadata statistics or throughput on an InfiniBand network.
Other similar tools include Ganglia and Nagios.

Additional user-level tools can be executed to get a basic understanding of what the application is doing.
For example, one can compare the capacity and node quotas before and after
job execution to check how much permanent data and how many files have been created.
Some tools work only for special file system or network types and need to be executed on client nodes where the application is running.

Additional tools of note are the Integrated Performance Monitoring for HPC\footnote{IPM, \url{https://github.com/nerscadmin/IPM}}, IOSIG\cite{yin2012boosting}, RIOT\cite{wright2012parallel}, ScalaIOTrace\cite{vijayakumar2009scalable}, and Linux blktrace.

\subsection{Further Administrative Tools}

Most preassembled parallel file system appliances provide a monitoring
system with a GUI. It displays various aspects of the servers such as load
and I/O activity for current and previous time slots.
Typically, it also shows a number of file-system-specific statistics.
In many cases, the monitoring systems are only accessible by system
administrators, and hence only a few HPC sites expose such information to their users.

However, statistics can be gathered from either each client node or the server nodes.
Correlation of a parallel application with the triggered activities on I/O
servers is non-trivial since these are shared.
With Lustre, \emph{jobstats}\,\cite{jobstats} provide a way to easily
collect I/O statistics of batch jobs with little overhead, for example, the number of open operations or the total amount of written data.
To activate \emph{jobstats}, a system administrator selects the environment
variable that holds the batch job ID on the client nodes. Lustre clients
send the content of this variable with the standard Lustre protocol to the
servers, and the servers sum up all I/O activity on this content. An example of how these statistics can be easily made available for users is described in \cite{lustre_mon_stats}.
The Lustre Monitoring Tool (LMT)~\cite{Uselton2009} is also available for
collecting server-side time series statistics from Lustre file systems, although data
produced with LMT does not distinguish traffic from different applications.

The first tool that combined actual client and server traces of I/O into one
view to show a comprehensive timeline of activities was PIOviz~\cite{ludwig2007analysis}.

\subsection{Tools for Unifying Trace Formats}

The diversity of file formats for profiles and tracing is approached by
tracing formats such as OTF~\cite{eschweiler2011open}, TAU, and EXTRAE by providing converters between the formats.
Since the formats differ slightly, sometimes some information is lost.
Besides such generic trace formats, various specialized trace formats exist.

Work is being done at the University of California, Santa Cruz, to produce a tool that provides field per record access to traces with the ability to convert them to another trace schema format on the fly.
By giving the ability to translate traces directly to other formats, there is the expectation that analysis will not be done ad hoc for different storage systems. This enables the creation of a standard set of tools for analyzing the behavior of parallel file systems without converting typically large traces.

While the database community has allowed programmable data presentation for some time \cite{smith2008guide}, through views and computed columns, these techniques have not worked their way into trace analysis, despite many system traces being stored in database formats, such as \texttt{Parquet}, a Spark SQL queryable columnar format \cite{armbrustspark2015}.
Presenting data in a uniform format, similar to a trace, is essential but
does not require the overhead of a database, since most analysis is done in place, usually in a time series format.

%We are aware of the availability of time series databases, but due to indexes only existing %on a single time value, they do not serve our purpose \cite{bonnet_towards_2001}. %Additionally, because the size and due to the fact these traces grow continually, we do not %want to constantly ingest this data into a database. For that reason, we have explored NoDB %systems \cite{alagiannis_nodb:_2012}, which allow for data to be queried in its raw format, %without defined schema.

The trace unification tool being developed borrows techniques from NoDB
systems as well as from the database community at large. Because of the
sequential nature and read-only analysis properties of trace analysis, our
aim is a way to enable low-latency translation from one format to another.
%We believe this is crucial to presenting a uniform format for trace analysis, instead creating a mechanism for schema translation (which we consider to be similar to computed columns in a full database engine that gives the tools necessary to present uniform versions of each trace without storing their information in a database or requiring views to be created before analysis.
%The goal of this tool is to create an information addressing system that allows for schemas to be translated as data is fetched from its base format with low overhead.
This approach can be used to unify traces based on the information they
share and allow for the information to be analyzed directly, either from a
storage system trace analysis tool that we plan to create or from an API
that we plan to add in future work.

This system ingests minimal information about each file in order to allow for field-per-record addressing, which is a good interface for analysis tools to access trace data.
This mechanism of trace addressing is used to create field and record
primitives, which serve as the basis for our schema translation language. By allowing for records and fields to be used as primitives, a simple mechanism for traces to be unified is provided.
This is achieved with a simple translation language based on s-expressions that allows for simple translation to occur, with more functionality coming in future work.
%Currently, our method for doing so is focused at the trace analysis community, though in future work we plan to provide additional functionality to allow for more complex translations to occur.

Other projects, such as TOKIO~\cite{lockwood2017umami,lockwood2018cug},
have explored the possibility of synthesizing normalized data on
demand via modular libraries rather than standardizing on an at-rest
trace format.  This approach is well-suited to integration of vendor
instrumentation tools that utilize proprietary formats or are otherwise
difficult to modify.  It also enables the integration of multiple data
sources simultaneously for holistic analysis and correlation.

\section{Example Studies}
\label{sec:examples}

This section gives examples about how the aforementioned tools can be used to identify performance issues.
First, we describe some performance issues that can arise.

System administrators typically try to identify users with the highest I/O activity because of the huge impact that can be realized by
helping such users reduce their I/O. Often users are not aware of what they are doing; for example, a user may forget to remove debugging
output, and the application will write many times more data than normal. Sometimes this is sufficient to crash the storage completely.

Many sites have different options for storing data: local SSDs on the cluster nodes, burst buffers, different parallel file systems, and so forth.
Frequently, application runtime can be improved by selecting an appropriate storage device.
For example, with the application OpenFoam~\cite{jasak2007openfoam},
one usually
can use local disks to store scratch data. Doing so increases the scalability of the application and reduces the load on the
central parallel file system, thus helping accelerate other applications.
While these optimizations are simple to apply,
users commonly store data in the wrong place and hence increase the load on the file system inadvertently.

With Lustre \emph{jobstats} and a simple perl script~\cite{lustre_mon_stats},  one can display all jobs with I/O activity
above a high-water mark, for example, jobs that have done more than 5 million open operations. Huge amounts of metadata operations should be
omitted because each operation requires communication with the responsible server. Lots of small read and write operations have a similar effect
and additionally might cause numerous slow seeks on the file system disks. Again, \emph{jobstats} can be used to find
jobs doing a huge number of read or write operations with a small average I/O size. For more detailed information about job I/O patterns
one would need to combine the file system metrics with application-level tracing from a tool such as Mistral.

Metadata operations are often a source of performance degradation. Common issues include trawling the file system looking for a file, which
results in lots of failed stat or open operations. Programs also commonly check the existence of a file before opening it,
producing an extra stat for every open.  A better approach would be to try to open the file and to fail gracefully if it is not there.
Users also can significantly slow metadata servers by opening a file every time it is written to. This usually happens
when the open operation has been placed inside a for loop.

In general, parallel file systems show better performance with large sequential read/write operations than with small random operations
(huge IOPS rates). Therefore, applications should read or write data in huge chunks. One area of HPC that is known for small I/O operations is the life-sciences industry\cite{ellexus_life} where applications often use one-byte reads and writes and
almost always use 32 kB reads and writes. Although small I/O operations may be justifiable when the algorithms employed in
mapping genomes lend themselves to the generation of lots of small files of around 4 kB, often the I/O operations are far smaller, leaving
room for improvement.

Another easy improvement with parallel file systems
can be achieved by setting the appropriate stripe count. If many files are used and if they are not accessed at the same time by many
tasks, a stripe count of 1 is appropriate. On the other hand, if files are shared by many tasks or if only a few tasks use huge files,
increasing the stripe count usually improves performance.

With MPI-IO, the MPI library, underlying libraries, and its adaption for the underlying file system can make a huge difference.
For example, some vendors provided MPI versions with activated Lustre support providing
further optimizations. With the main optimization each collective buffering
node collects the I/O for one OST, in other words, small
I/O buffers are collected into bigger buffers, and there are no locking conflicts between different nodes which try to access the same
area of data. Thus, selecting the right number of collective buffering nodes for a given application and problem size might
improve its runtime.

%%%%%%%%%%%%%%%%%%%%%%%%%%%%%%% BEGIN Example PIOM-MP %%%%%%%%%%%%%%%%%%%%%%%%%%%%%%%%%%%%%%%%%%%
\subsection{I/O Performance Analysis at the Application level}
The interaction between the I/O system and the application pattern can report poor
performance in some HPC systems. In order to identify the root cause of the problem,
the I/O pattern along the I/O path needs to be analyzed. Because the I/O
system spans between user (compute nodes) and administrator domains, sometimes it is very difficult to coordinate work to find a possible solution.

On the one hand, administrators
have several monitoring tools and commands to control, identify, and solve problems. Usually,
these tools are not accessible by the users, however. Problems on this level
are related mainly to parallel file systems, the network, and storage devices.

On the other hand, users have several performance analysis tools that,
depending on the problem,
can be used as profiling or tracing to identify or analyze the I/O.
At this level, I/O issues are related mainly to application pattern and I/O libraries.
The user needs to understand the impact that a specific access pattern can have on the performance.
Achieving such understanding is not simple, however, because a real
application can use several files and the user must
evaluate each file to weigh the impact of each one, which not always is related
directly with file size. An appropriate tool to start with for I/O performance analysis
is Darshan, which provides I/O time and throughput for each file opened by a parallel application.

\begin{figure}[bt!]
  \begin{subfigure}{0.47\textwidth}
  \subcaption{I/O Problem}
  \includegraphics[width=\textwidth]{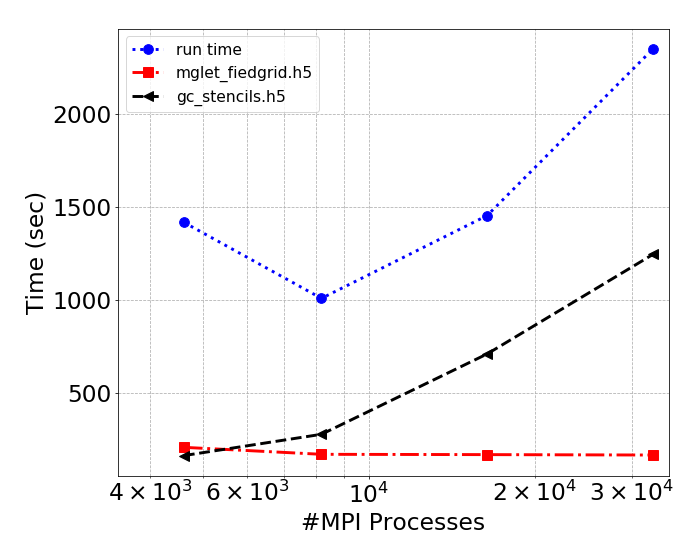}
  \label{fig:mglet-io-issues:a}
  \end{subfigure}
  \quad
  \begin{subfigure}{0.48\textwidth}
  \subcaption{\texttt{gc\_stencils.h5} file offset}
    \includegraphics[width=\textwidth]{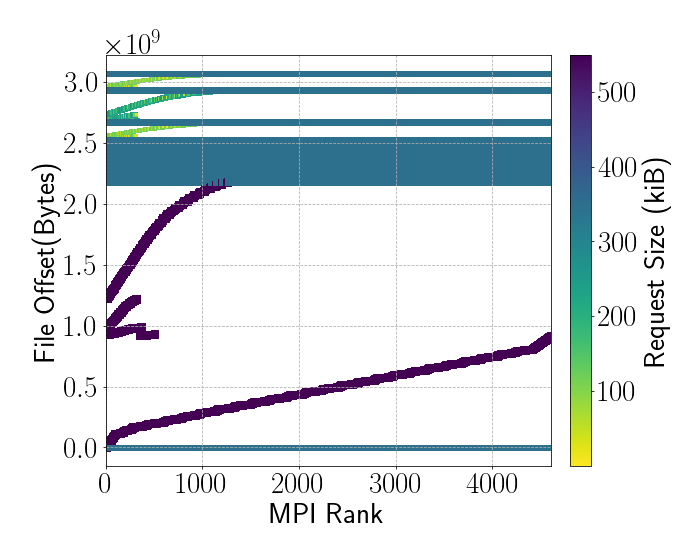}
    \label{fig:mglet-io-issues:b}
  \end{subfigure}
  \begin{subfigure}{0.46\textwidth}
  \subcaption{The \texttt{stencils} file temporal pattern}
  \includegraphics[width=\textwidth]{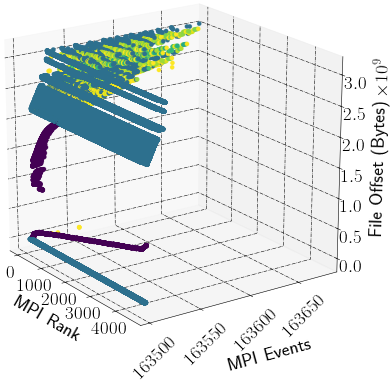}
   \label{fig:mglet-io-issues:c}
  \end{subfigure}
    \begin{subfigure}{0.53\textwidth}
  \subcaption{Timestamp for the \texttt{stencils} file }
  \includegraphics[width=\textwidth]{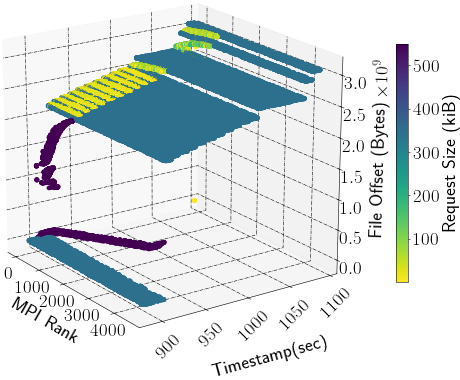}
   \label{fig:mglet-io-issues:d}
  \end{subfigure}
  \vspace{-5mm}
  \caption{Example of I/O performance analysis. (a) Identifying the file with more impact on run
  time using Darshan. (b) Analyzing the I/O pattern with a focus on the \texttt{gc\_stencils.h5} file
  offset for 4608 MPI processes using PIOM-MP. (c) The \texttt{gc\_stencils.h5} file temporal pattern based on the application logical using MPI events as ticks. (d) The \texttt{gc\_stencils.h5} file temporal pattern using system timestamps.}
    \label{fig:mglet-io-issues}
\end{figure}

To illustrate this, \Cref{fig:mglet-io-issues:a} shows the I/O time and the relation with total time for
a strong scaling case that was obtained using Darshan. Two files are selected for analyzing the
application scalability: (1) the \texttt{mglet\_fieldgrid.h5} (red line) is the file that the user considers
more important for the application because move a total of 400GiB, which after
optimizations shows a scalable behavior; and (2) a small file that corresponds to ghost cells
(\texttt{gc\_stencils.h5} in black line) that is avoiding scaling to the application.
\Cref{fig:mglet-io-issues:b} shows the offset for the ghost file, whose size is
2.9 G. Concurrent accesses can be observed in \Cref{fig:mglet-io-issues:c} for all the MPI processes
in lowest and highest offsets, which serialize the I/O as can be seen in \Cref{fig:mglet-io-issues:d}.
The application presents an I/O imbalance observed for the first 1024 MPI processes. The I/O pattern represented
in \Cref{fig:mglet-io-issues:b} is similar for the different cases showed in \Cref{fig:mglet-io-issues:a} in which
there are several rewriting operations that are moving more data into the file system as the number of MPI
processes increases. In this case, one must redesign the I/O pattern in
order to remove the problem.

%%%%%%%%%%%%%%%%%%%%%%%%%%%%%%% END Example PIOM-MP %%%%%%%%%%%%%%%%%%%%%%%%%%%%%%%%%%%%%%%%%%%

\subsection{Online Monitoring}
\begin{figure}[bt!]
	\centering
	\includegraphics[width=0.9\textwidth]{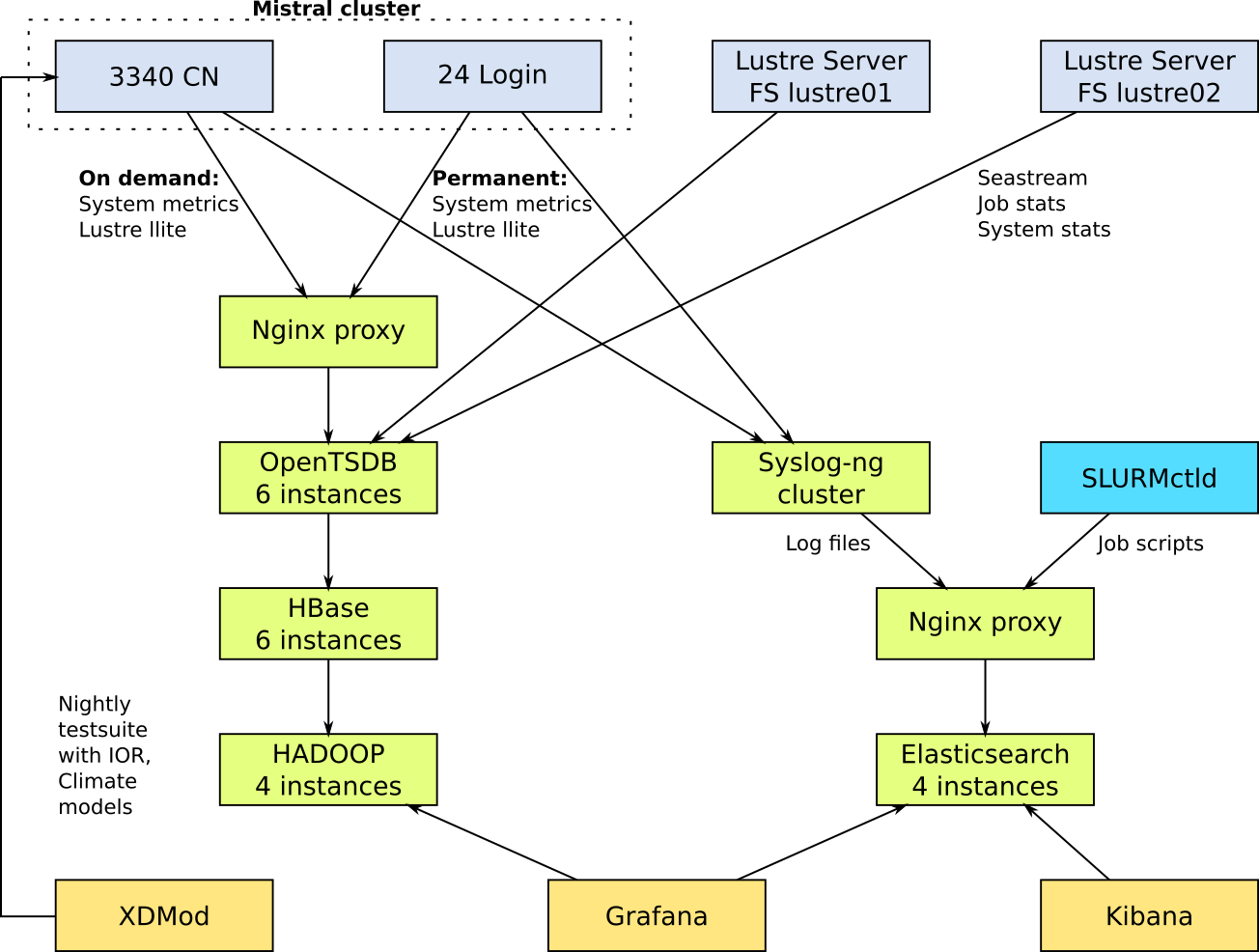}
	\caption{DKRZ-monitoring}
	\label{dkrz-mon}
\end{figure}

DKRZ maintains a monitoring system that gathers various statistics from 3,340 client nodes, 24 login nodes, and Lustre servers.
The monitoring system is realized mainly by open source components such as Grafana, OpenTSDB,\footnote{Consists of various additional components from the Hadoop Stack}, and Elasticseach but also includes a self-developed data collector.
Additionally, the monitoring system obtains various information from the Slurm workload manager.
A schematic overview is provided in \Cref{dkrz-mon}.

\begin{figure}[bt!]
	\centering
	\includegraphics[width=0.9\textwidth]{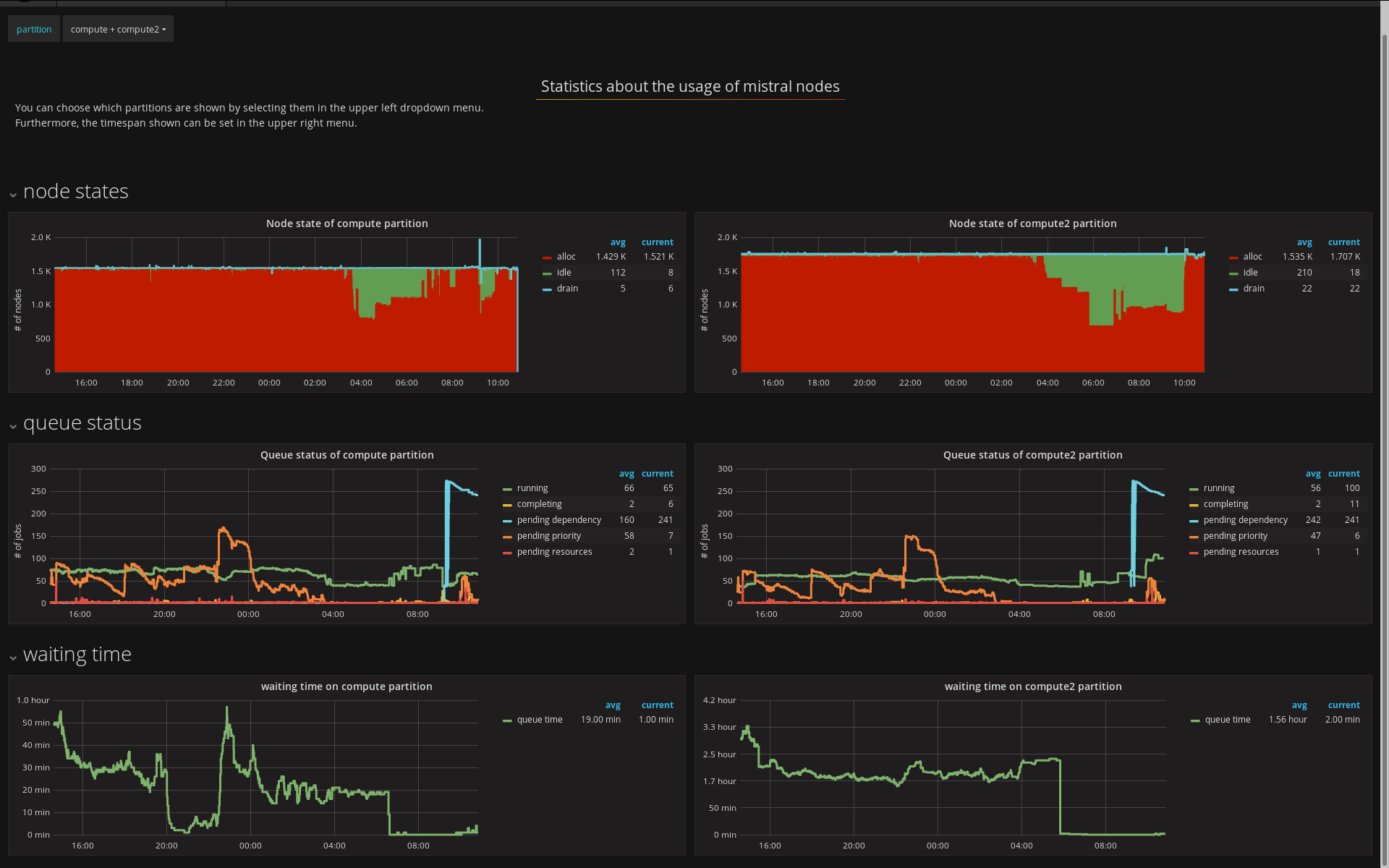}
	\caption{Statistics about the usage of Mistral nodes}
	\label{dkrz-mon-slurm}
\end{figure}

% Visualization
The data is aggregated and visualized by a Grafana web interface, which is available to all DKRZ users.
The information is structured in three sections: login nodes, user jobs, and queue statistics.

% Login
In the first place, a monitoring service gives the users an overview of the current state of the system, namely, the current load of login nodes and number of used nodes on Slurm partitions.
For each single machine, a detail view also provides information about system load, memory consumption, and Lustre statistics, as well as
historical data.
% Job
Job monitoring is enabled by default in a coarse-grained mode but can be modified by a Slurm parameter.
When enabled, the monitoring system gathers information about CPU usage and frequency, memory consumption, Lustre throughput, and network traffic for each client node.
% Queue
Statistics about the usage of \emph{Mistral} nodes (\Cref{dkrz-mon-slurm}) show the current state and history of node allocation, queue status, and waiting time of both Slurm partitions.
Additionally, DRKZ runs xdmod on the client nodes for viewing historical job information as well as real-time scientific application profiling.

\subsection{Online Monitoring with LLview}

In 2004 the J\"ulich Supercomputing Centre (JSC) began to develop
LLview~\cite{llview-web}, an interactive graphical tool to monitor job
scheduling for various resource managers such as SLURM or
LoadLeveler. In its stand-alone and web-based clients LLview provides
information about currently running jobs and how they are mapped on the
individual nodes of an HPC cluster, using an intuitive visual
representation of the underlying cluster configuration.  In addition,
it includes statistical information on completed jobs as well as
a prediction for scheduled future jobs~\cite{llview-jufo}.  LLview is
client-server based. Internally it uses an XML schema named LML
\cite{LLview-LML} as an extensible abstraction layer to encode monitor
data. The data itself is obtained from various server and application
interfaces and subsequently transferred to the clients of LLview. LML
is a simple, extensible, and independent description of node- and
job-related information. A subset of the LLview dashboard elements are
also integrated in Parallel Tools Platform, an extension to
Eclipse to develop, run, and debug parallel programs on a remote HPC
system \cite{eclispe-PTP}.

Recently, a number of extensions to LLview have been designed and
implemented that allow users to acquire, store, and display job performance
metrics such as node CPU load, memory usage, interconnect activity, and
I/O activity. In practice at JSC these have greatly facilitated
the analysis of HPC codes that either show unsatisfactory performance
or behave incorrectly. With regard to metrics about I/O, the subject of
this paper, LLview uses performance data generated with the GPFS
mmpmon command~\cite{GPFS-mmpmon} within the framework of the IBM
Spectrum Scale (GPFS) file system. At JSC, all GPFS clients running on
cluster compute nodes are configured to frequently write (every 1--2
minutes) such mmpmon data to a log file that itself resides in a local
GPFS files system. The LLview server components collect the I/O
performance data (e.g., number of bytes written/read, number of
open/close calls) from these files and match it to user jobs that are
running on these nodes.  This mapping is possible because
node sharing is not applied on JSC systems. The job-based data is then
stored in an internal database, from which individual performance
reports are generated.

Since the minimum update frequency of LLview is not more than once a
minute, JSC has integrated this to help users monitor job behavior on
a coarse level and to help detect performance issues while jobs are
running on the system. Within a web portal (\emph{LLview jobreports})
access is provided to timeline-based performance data of current and recently
finished jobs on the system, ranging two weeks into the
past. In addition to interactive charts displaying the various
performance data, the portal provides PDF reports for all jobs,
which can be collected and archived by users to document their
production runs. These extensions of LLview have proven useful as
a first step of performance analysis and code tuning, thanks to the
integrated coarse view on the most essential performance metrics of
user jobs. In this way it greatly facilitates the detection of
situations where deeper performance analysis and tuning should be
started with additional tools such as Darshan or SIOX (\Cref{sec:tools}).

\section{Challenges in Analyzing I/O}
\label{sec:issues}
The following list represents issues that we see as a community that must be overcome to ensure effective and efficient monitoring in future HPC systems.

%- things we are doing well (things that are well covered by the field in general, kind of a counterpoint to the first bullet)

\begin{itemize}
\item \textbf{Ability to monitor I/O at all levels}: The involvement of many layers in the I/O stack that are not instrumented limits the ability to identify issues. Today, parallel file systems typically have disks, storage controllers, storage servers, and clients. All are connected with different network types. Hardware and firmware or software can cause performance degradation at each layer. Above the parallel file systems are MPI-IO, MPI libraries, and upper-level libraries such as HDF5.
In many cases administrative power and deep knowledge are required in order to capture data from certain layers.
In an ideal case, one should be able to easily investigate the I/O at each level.

\item \textbf{Adaption of tools to new programming models}: The HPC application community has long been (and still is) dominated by
MPI simulations written in Fortran or C.  However, other paradigms such as deep learning and big data use their own programming
models, languages, and runtime environments.  Parallel I/O analysis tools must adapt to keep pace as these models are more widely adopted in HPC.

\item \textbf{Adaption to new hardware technologies}: Analyzing parallel I/O at the moment basically means analyzing I/O access to a parallel file system, but a number of emerging technologies call for a broader interpretation of I/O.  Future systems may include nonvolatile memory (which may be accessed as a memory device or a file system) and multiple tiers of storage that hold the simulation working
set, campaign data, or archival data. The performance of each subsystem is best interpreted in the context of the complete storage hierarchy.
These new technologies must provide means to systematically report performance data for the analysis.

\item \textbf{Integrating analysis data}:
Data analysis of performance data is often an afterthought, done after the application was executed (post mortem).
Projects such as TOKIO~\footnote{http://www.nersc.gov/research-and-development/storage-and-i-o-technologies/tokio/} are working toward integrating data from multiple
components of the storage system, but this remains a pressing problem.  Each storage system component is typically designed by a
specialized vendor and is serviced by its own instrumentation framework.
Aligning and synthesizing data
from job schedulers, applications, file systems, storage devices, and network devices remain a significant challenge.

\item \textbf{Feature replication and sustainability}: Many of the technologies included in this survey serve overlapping purposes and
implement their own variation of tasks such as function interception, time series indexing, and visualization.
Increased standardization and code sharing could potentially reduce the engineering effort involved in each project while at the same time increasing portability.

\item \textbf{Performance assessment}: Monitoring and recording performance data are part of the process.
However, it is often hard for experts to assess whether the observed behavior and performance is acceptable or not.
End users typically do not understand what a runtime of X\,seconds or an achieved performance of Y\,GiB/s means.
Guidance for assessing the quality is a necessary step.

\item \textbf{Correlation of I/O monitoring with other metrics}: In
  parallel applications it is often essential to observe more than only one
  performance metric such as I/O activity in an isolated manner. Rather,
  bringing I/O activity in correlation with other metrics such as CPU
  activity, memory usage, and interconnect usage is needed in order to identify
  root causes of faulty behavior and to distinguish between causes and
  effects.
  %The above mentioned job reporting tool of LLview (or DKRZ's Grafana) is an example for an entry point to such an analysis without the need for   any instrumentation preparation of applications.

\item \textbf{User guidance}: A gap remains between what can be learned from expert I/O analysis and what can be readily adapted by
end users.
The data produced by today's I/O analysis tools requires significant expertise to interpret, and little guidance is provided as
to what I/O performance a user should expect in a given application scenario.

\item \textbf{Small developer community}: I/O tools for HPC suffer from the fact that the HPC community is small and I/O is often neglected.
%GAIL - I don't understand this: "as it is as a supportive necessity for computing."
Massive amounts of traces are being generated from user applications and other sources outside the HPC community but not covered by tools.
We have to make the base of tools (at least) useful outside the HPC community, for example, to CERN~\cite{peters_eos_2015, peters_exabyte_2011}, in order to speed the development.
\end{itemize}

We believe that the \textbf{standardization of monitoring APIs and infrastructure} is a necessary step facing the overall community.
At the moment no standard API exists to ease the reporting of data; instead, vendors and tool developers build their own solution.
A standard would be beneficial to several issues mentioned in this list,
and might help encourage hardware and software developers to provide the necessary data.
Indeed, we must standardize our approaches to system analysis, both to allow comparison between like systems and to avoid the programming overhead of repeatedly creating the same analytics code again and again. By creating a standardized method of analyzing storage systems, as well as other system traces, we hope to light the path ahead to exascale and help computing centers deal with massive growth and data-driven analysis.

Additionally, by creating a general-purpose method of trace analysis, we can provide optimizations and features that would be too costly to develop for individual tools or systems.

\section{Conclusions}
\label{sec:conclusion}

In this paper, we have provided an overview of the state-of-the-art in I/O monitoring tools, motivated by our own involvement in the development of the described tools.
We then gave examples of how monitoring can be used to reveal I/O issues; because of space limitations, this list is intended as an appetizer for the reader to study the tools and mentioned papers further.
Furthermore, we discussed challenges we see as a community that are currently not well addressed and must be resolved further in future research, development, and engineering.
One step forward for the community is to work on standardization.

\section*{Acknowledgments}
PIOM-MP is a work partially supported by the MICINN/MINECO Spain under contracts
TIN2014-53172-P and TIN2017-84875-P.
This material is based in part on work supported by the U.S. Department of Energy, Office of Science, under contract DE-AC02-06CH11357.
We thank Felicity Pitchers for the proofreading.

\bibliography{bibliography,UniformTrace}
\end{document}